\documentclass[twocolumn,showpacs,amsmath,amssymb,10pt,aps]{revtex4}
\usepackage{graphicx,color}
\usepackage{amsmath}
\usepackage{amssymb}
\usepackage{bbm}
\usepackage{color}

\usepackage[utf8]{inputenc}

\usepackage{amsfonts}
\usepackage{bm}
\newcommand{\rr}{\mathbb{R}}

\newcommand{\ud}{\mathrm{d}}

\newcommand{{\Cd}}{{\mathbb{C}^d}}

\def\oper{{\mathchoice{\rm 1\mskip-4mu l}{\rm 1\mskip-4mu l}
{\rm 1\mskip-4.5mu l}{\rm 1\mskip-5mu l}}}
\def\<{\langle}
\def\>{\rangle}
\newtheorem{Theorem}{Theorem}
\newtheorem{Lemma}{Lemma}

\newtheorem{property}{Property}
\newtheorem{Corollary}{Corollary}

\newtheorem{Remark}{Remark}
\newtheorem{Example}{Example}
\newtheorem{Proposition}{Proposition}

\newcommand{\beq}{\begin{equation}}
\newcommand{\eeq}{\end{equation}}
\newcommand{\bear}{\begin{eqnarray}}
\newcommand{\ear}{\end{eqnarray}}
\newcommand{\bdm}{\begin{displaymath}}
\newcommand{\edm}{\end{displaymath}}

\newcommand{\lhs}{\mathrm{LHS}}
\newcommand{\rhs}{\mathrm{RHS}}

\begin{document}
\title{\textbf{Admissible memory kernels for random unitary qubit evolution}}
\author{Filip A. Wudarski,  Pawe\l\ Nale\.zyty, Gniewomir Sarbicki and Dariusz Chru\'sci\'nski}
\affiliation{ Institute of Physics, Faculty of Physics, Astronomy and Informatics \\  Nicolaus Copernicus University,
Grudzi\c{a}dzka 5/7, 87--100 Toru\'n, Poland}



\begin{abstract}
We analyze random unitary evolution of a qubit within memory kernel approach. We provide sufficient conditions which guarantee that the corresponding memory kernel generates physically legitimate quantum evolution. Interestingly, we are able to recover several well-known examples and to generate new classes of nontrivial qubit evolution. Surprisingly, it turns out that a class of quantum evolutions with memory kernel generated by our approach gives rise to the vanishing of a non-Markovianity measure based on the distinguishability  of quantum states.
\end{abstract}

\maketitle

\section{Introduction}

Dynamics of open quantum systems plays an important  role in the analysis of various phenomena like dissipation, decoherence and dephasing  \cite{Breuer,Weiss}. The usual approach to the dynamics of an open quantum
system consists of applying the Born-Markov approximation \cite{Breuer}
which leads to a local master equation for the Markovian semigroup
\begin{equation}\label{ME1}
  \dot{\rho}_t =L [\rho_t]\ ,
\end{equation}
where $\rho_t$ is the density matrix of the investigated system  and $L$
is the time-independent generator of the dynamical semigroup defined as follows
\beq\label{Lin}
L[\rho]=-i[H,\rho]+\frac{1}{2}\sum_\alpha\left([V_\alpha,\rho V_\alpha^\dag]+[V_\alpha\rho,V_\alpha^\dag]\right).
\eeq
Here $H$ denotes the effective system Hamiltonian, and $V_\alpha$ represent noise operators \cite{GKS,L}. We call (\ref{Lin}) the GKSL form (Gorini-Kossakowski-Sudarshan-Lindblad).
The solution of (\ref{ME1}) defines the Markovian semigroup
\begin{equation}\label{}
  \rho_t = \Lambda_t[\rho] = e^{tL} \rho\ ,
\end{equation}
where $\rho$ is an initial state. The dynamical map $\Lambda_t = e^{tL}$ is completely positive and trace-preserving (CPTP) \cite{Breuer,GKS,L,Alicki}. The Born-Markov approximation assumes weak interaction and a separation of time scales between the system and its environment. Such approach works perfectly well for many quantum optical systems \cite{Gardiner,Plenio,Car}. When the above assumption is no longer valid the description based on (\ref{ME1}) is not satisfactory.  Recent technological progress and modern laboratory techniques call for a more refined approach which takes into account memory effects  completely neglected in the description based on Markovian semigroups. In recent years we observed an intense research activity in the field of non-Markovian quantum evolution (see the recent review \cite{RIVAS}, a collection of articles in \cite{JPB} and a recent comparative analysis in \cite{Addis}).

There are basically two approaches which generalize the standard Markovian master equation (\ref{ME1}): time-local approach replaces $L$ by a time-dependent generator $L_t$. Interestingly, if for all $t$ the time-dependent generator has the standard GKSL form (\ref{L}), then $\Lambda_t = \mathcal{T} \exp(\int_0^t L_u du)$ defines the so-called divisible dynamical map \cite{Wolf-Isert,RHP} which is often considered as the generalization of Markovianity (see \cite{Sab} for a generalization of the notion of divisibility). The second approach is based on the nonlocal Nakajima-Zwanzig equation \cite{NZ} (see also \cite{NZ-inni})
\begin{equation}\label{NZ}
  \dot{\rho}_t = \int_0^t K_{t-\tau} \rho_\tau d\tau ,
\end{equation}
in which quantum memory effects are taken into account
through the introduction of a memory kernel $K_t$. This means that the rate of change of the state $\rho_t$ at time $t$ depends
on its history (starting at t = 0). The Markovian master Eq. (1)
is reobtained when $K_t = 2\delta(t)L$. The time-dependent kernel
is usually referred to as the generator of the non-Markovian
master equation. Equation (\ref{NZ}) applies to a variety of situations (see eg. \cite{Fabio}).
Because of the convolution structure of (\ref{NZ}) the time-local approach is often called time-convolutionless \cite{Breuer,Hangii,PRL-2010}. The structure and the properties of (\ref{NZ}) were carefully analyzed in \cite{bar,Budini,Wilke,B-V,Wod,Maniscalco,Lidar,AK,EPL,Bassano}. In particular the generalization of Markovian evolution to the so-called semi-Markov was investigated within the memory kernel approach by Budini \cite{Budini} and Breuer and Vacchini \cite{B-V} (see also discussion in \cite{EPL}).

In a present article we study random unitary evolution of a qubit within the memory kernel approach. In particular we address the following problem: what is the structure of the corresponding memory kernel $K_t$ which leads to the legitimate CPTP dynamical map $\Lambda_t$.
The article has the following structure: in Section II we recall basic facts about random unitary evolutions and in Section III we formulate the sufficient condition for $K_t$ to guarantee legitimate physical evolutions. In Section \ref{MAR} we examine the issue of Markovianity. Surprisingly, it turns out that a subclass of quantum evolutions with memory kernel generated by our approach gives rise to the vanishing of a non-Markovianity measure based on the distinguishability  of quantum states \cite{BLP}.  Section V illustrates our approach with several examples. Final conclusions are collected in Section VI.


\section{Random unitary qubit evolution}

A quantum channel $\mathcal{E} : \mathcal{B}(\mathcal{H}) \rightarrow \mathcal{B}(\mathcal{H})$ is called random unitary \cite{Ad} if its Kraus representation is given by
\begin{equation}\label{}
  \mathcal{E}[X] = \sum_k p_k\, U_k XU_k^\dagger\ ,
\end{equation}
where $U_k$ is a collection of unitary operators and $\{p_k\}$ stands for a probability distribution. In this article we consider a random unitary dynamical map $\Lambda_t$ defined by 
\begin{equation}\label{RU}
  \Lambda_t[\rho] = \sum_{\alpha=0}^3 p_\alpha(t)\, \sigma_\alpha \rho \sigma_\alpha\ ,
\end{equation}
where  $\sigma_\alpha$ are Pauli matrices with $\sigma_0 = \mathbb{I}_2$ \cite{Karol}.  Initial condition $\Lambda_{t=0} = \oper$ implies $p_\alpha(0) = \delta_{\alpha 0}$. Recently a time-local description based on the following master equation was analyzed \cite{Erika,PLA}
\beq
\dot{\Lambda}_t=L_t\Lambda_t ,
\eeq
where $L_t$ is a time-local generator defined by
\beq\label{L}
L_t[\rho]=\sum_{k=1}^3\gamma_k(t)\left(\sigma_k \rho \sigma_k-\rho\right),
\eeq
with time-dependent decoherence rates $\gamma_k(t)$. One asks the following question: what are the conditions for $\gamma_k(t)$ which guarantee that the solution $\Lambda_t = \exp(\int_0^t L_\tau d\tau)$ provides a legitimate dynamical map? Note, that the solution defines a random unitary evolution with $p_\alpha(t)$ given by
\begin{equation}\label{p-H}
  p_\alpha(t) = \frac 14 \sum_{\beta=0}^3 H_{\alpha\beta} \lambda_\beta(t)\ ,
\end{equation}
where $H_{\alpha\beta}$ is the Hadamard matrix
\beq\label{had}
	H=\left(\begin{array}{rrrr}1 & 1 & 1 & 1 \\1 & 1 & -1 & -1 \\1 & -1 & 1 & -1 \\1 & -1 & -1 & 1\end{array}\right),
\eeq
and $\lambda_\beta(t)$ are time-dependent eigenvalues of $\Lambda_t$
\begin{equation}\label{}
  \Lambda_t[\sigma_\alpha] = \lambda_\alpha(t)\sigma_\alpha\ ,
\end{equation}
which read as follows: $\Lambda_0(t)=1$ and
\begin{eqnarray}\label{}
  \lambda_1(t)& =& \exp(-2[\Gamma_2(t) + \Gamma_3(t)]), \nonumber\\
   \lambda_2(t)& =& \exp(-2[\Gamma_1(t) + \Gamma_3(t)]) , \\
    \lambda_3(t)& =& \exp(-2[\Gamma_1(t) + \Gamma_2(t)]), \nonumber
\end{eqnarray}
with $\Gamma_k(t) = \int_0^t \gamma_k(\tau)d\tau$. Now, the map (\ref{RU}) is CP iff $p_\alpha(t) \geq 0$, which is equivalent to the following set of conditions for $\lambda$s  \cite{Erika,PLA}:
\begin{equation}\label{lambda-0}
  1 +  \lambda_1(t) + \lambda_2(t) +  \lambda_3(t) \geq 0 \ ,
\end{equation}
and
\begin{eqnarray}\label{lambda-123}
  \lambda_1(t) + \lambda_2(t) &\leq& 1 + \lambda_3(t), \nonumber\\
  \lambda_3(t) + \lambda_1(t) &\leq &1 + \lambda_2(t), \\
  \lambda_2(t) + \lambda_3(t) &\leq& 1 + \lambda_1(t). \nonumber
\end{eqnarray}

\section{Construction of legitimate memory kernels}

In this article we analyze the nonlocal description based on the following memory kernel equation
\beq	\label{splot}
	\dot{\Lambda}_t = \int\limits_{0}^{t} K_{t-\tau} \Lambda_\tau \ud \tau\ ,
\eeq
with
\begin{equation}\label{}
  K_t[\rho] = \sum_{i=1}^3  k_i(t)\left(\sigma_i \rho \sigma_i-\rho\right),
\end{equation}
where $k_i(t)\ (i=1,2,3)$ represent nontrivial memory effects. Note, that equation (\ref{splot}) considerably simplifies after performing the Laplace transform
\beq
	\tilde{\Lambda}_s=\frac{1}{s-\tilde{K}_s} ,
\eeq
where $\tilde{\Lambda}_s := \int_0^\infty e^{-st} \Lambda_t dt$ and similarly for $\tilde{K}_s$.
The question we address is: {\em what are the conditions for $k_i(t)$ which guarantee that the solution $\Lambda_t$ provides a legitimate dynamical map}?

Denoting by $\kappa_\alpha(t)$ the eigenvalues of $K_t$,
\begin{equation}\label{}
  K_t[\sigma_\alpha] = \kappa_\alpha(t)\sigma_\alpha\ ,
\end{equation}
equation (\ref{splot}) gives rise to the following set of equations:
\beq	\label{splot_skalar}
	\dot{\lambda}_i(t)=\int_0^t\kappa_i(t-\tau)\lambda_i(\tau)\ud\tau , \ \ \ i=1,2,3.
\eeq
Note, that $\kappa_0(t) = 0$ and hence $\lambda_0(t) = 1 = {\rm const}$. In terms of the Laplace transforms $\tilde{\lambda}_i(s)$ and
$\tilde{\kappa}_i(s)$ one finds
\begin{equation}\label{l-k}
  \tilde{\lambda}_i(s) = \frac{1}{ s- \tilde{\kappa}_i(s)  } \ .
\end{equation}
In terms of $\tilde{\lambda}_i(s)$ conditions (\ref{lambda-0})--(\ref{lambda-123}) may be equivalently reformulated as follows:
\begin{equation}\label{s-0}
   \frac 1s + \tilde{\lambda}_1(s) + \tilde{\lambda}_1(s) + \tilde{\lambda}_2(s) \ \ \ \mbox{is CM} \ ,
\end{equation}
and
\begin{eqnarray} \label{s-123}
  \frac 1s + \tilde{\lambda}_3(s) - \tilde{\lambda}_1(s) - \tilde{\lambda}_2(s)  && \ \mbox{is CM} \ , \nonumber \\
  \frac 1s + \tilde{\lambda}_2(s) - \tilde{\lambda}_1(s) - \tilde{\lambda}_3(s)  && \ \mbox{is CM} \ , \\
  \frac 1s + \tilde{\lambda}_1(s) - \tilde{\lambda}_3(s) - \tilde{\lambda}_2(s)  && \ \mbox{is CM} \ , \nonumber
\end{eqnarray}
where CM stands for a completely monotone function \cite{CM}, i.e. a smooth function $f:\ [0,\infty)\rightarrow\rr$ satisfying the  condition
\beq	\label{cm}
	(-1)^n\frac{\ud^n}{\ud s^n}f(s)\ge0, \quad s\ge0,\ \ n=0,1,2,\ldots
\eeq
The equivalence of (\ref{lambda-123}) and (\ref{s-123}) results from the following

\begin{Theorem}[Bernstein's Theorem]
	A function $f\colon[0,\infty)\to\rr$ is completely monotone on $[0,\infty)$  if and only if it is a Laplace transform of a finite non-negative Borel measure $\mu$ on $[0,\infty)$, i.e. $f$ is of the form
	\beq
		f(s)=\int_0^\infty e^{-st}\ud\mu(t).
	\eeq
\end{Theorem}
Note that the initial condition $p_0(0)=1$ and $p_k(0)=0$ for $k=1,2,3$ is equivalent to $\lambda_k(0)=1$ due to
\begin{Theorem}[Initial Value Theorem]
	Let $\tilde{f}(s)$ be the Laplace transform of $f(t)$. Then the following relation is true:
	\beq
		\lim_{t \rightarrow 0} f(t) = \lim_{s \rightarrow \infty} s\tilde{f}(s)
	\eeq
\end{Theorem}
it is equivalent to
\begin{equation}\label{ini}
  \lim_{s \rightarrow \infty} s\tilde{\lambda}_k(s) = 1 \ ,
\end{equation}
for $k=1,2,3$. This way we have proved

\begin{Theorem} The map $\tilde{\Lambda}_s$ represented by the following spectral decomposition
\begin{equation}\label{}
  \tilde{\Lambda}_s[\rho] = \frac 12 \sum_{\alpha=0}^3  \tilde{\lambda}_\alpha(s) \sigma_\alpha {\rm tr}[\sigma_\alpha \rho] \ ,
\end{equation}
with $\tilde{\lambda}_0(s) = 1/s$, defines the Laplace transform of a legitimate map $\Lambda_t$ if and only if conditions (\ref{s-0}), (\ref{s-123}) and (\ref{ini}) are satisfied.
\end{Theorem}
It is worth  emphasising  that there are few analytical tools for dealing with CM functions, which is due to the fact that an infinite set of conditions \eqref{cm} must be verified. Nevertheless, we found an important class of CM functions giving rise to CPTP dynamics with a straightforward interpretation. To present them, let us first observe that  CM functions have the following two properties, which will not be proved:
\begin{property}	\label{property1}
	Let $f$ and $g$ be arbitrary completely monotone functions. Then
	%
	\begin{enumerate}
		\item $ f\cdot g$ is CM,
		\item  $\alpha f+\beta g$ is CM for any ${\alpha,\beta>0}$,
	\end{enumerate}
\end{property}
\begin{property}	\label{property2}
	If $s_0 \ge 0$ then $\frac{1}{s+s_0}$ is CM.
\end{property}

We are now ready to prove our main result:
\begin{Theorem}\label{TH}
Let $W(s)$ be a function such  that $ \frac 1s \frac{1}{W(s)}$
is CM. Then  the functions
\begin{equation}\label{kapcm}
\tilde{\kappa}_k(s)=-\frac{s}{a_k W(s)-1},\quad k=1,2,3,
\end{equation}
with $a_1,a_2,a_3 > 0$ such that
\begin{equation}\label{aaa-0}
   \frac 1s \left( 4 -  \frac{1}{W(s)} \left[ \frac{1}{a_1} + \frac{1}{a_2} + \frac{1}{a_3}\right] \right)   \  \  \ \mbox{is CM} \ ,
\end{equation}
and
\begin{eqnarray} \label{aaa}
  \frac{1}{a_1} + \frac{1}{a_2}  &\geq& \frac{1}{a_3} \ , \nonumber \\
  \frac{1}{a_2} + \frac{1}{a_3}  &\geq& \frac{1}{a_1} \ , \\
  \frac{1}{a_3} + \frac{1}{a_1}  &\geq& \frac{1}{a_2} \ , \nonumber
\end{eqnarray}
define a legitimate memory kernel
\begin{equation}\label{}
  \tilde{K}_s[\rho] = \frac 12 \sum_{k=1}^3  \tilde{\kappa}_\alpha(s) \sigma_k {\rm tr}[\sigma_k \rho] \ ,
\end{equation}
i.e. the corresponding $\tilde{\lambda}_k(s)$ satisfy (\ref{s-0})--(\ref{s-123}) and (\ref{ini}).
\end{Theorem}	
Proof: note that formula (\ref{kapcm}) implies
\begin{equation}\label{}
  \tilde{\lambda}_k(s) = \frac 1s \left( 1 - \frac{1}{a_kW(s)} \right)\ ,
\end{equation}
and hence
\begin{eqnarray}\label{}
 &&  \frac 1s + \tilde{\lambda}_3(s) - \tilde{\lambda}_1(s) - \tilde{\lambda}_2(s)   \nonumber \\
 &=& \frac 1s \frac{1}{W(s)} \left(
   \frac{1}{a_1} + \frac{1}{a_2}  - \frac{1}{a_3} \right),
\end{eqnarray}
which proves that $\frac 1s + \tilde{\lambda}_3(s) - \tilde{\lambda}_1(s) - \tilde{\lambda}_2(s) $ is CM due to the fact that $\frac 1s \frac{1}{W(s)}$ is CM. Similarly one proves the remaining conditions (\ref{lambda-123}). \hfill $\Box$

Note, that since $\frac 1s \frac{1}{W(s)} $ is CM, hence, due to the Bernstein theorem, it is the Laplace transform of a positive function. Hence
\begin{equation}\label{}
  W(s) = \frac{1}{\tilde{f}(s)} \ ,
\end{equation}
where $ \tilde{f}(s)$ is the Laplace transform of $f(t)$ satisfying $\int_0^t f(\tau)d\tau \geq 0$ for all $t\geq 0$. One finds
\begin{equation}\label{tilde-kappa}
  \tilde{\kappa}_k(s) = \frac{-s \tilde{f}(s)}{a_k - \tilde{f}(s)} \ .
\end{equation}
Note, that condition (\ref{aaa-0}) implies
\begin{equation}\label{aaa-0n}
  \left( \frac{1}{a_1} + \frac{1}{a_2} + \frac{1}{a_3}\right) \int_0^t f(\tau)d\tau \leq 4 \ .
\end{equation}
Hence to summarize: our class is characterized by a single function $f(t)$ and three numbers $a_1,a_2,a_3 > 0$
such that $F(t) = \int_0^t f(\tau)d\tau \geq 0$ and conditions (\ref{aaa}) and (\ref{aaa-0n}) hold.
One finds for $p_\alpha(t)$:
\begin{eqnarray} \label{pF}
  p_1(t) &=& \frac 14  \left( \frac{1}{a_2} + \frac{1}{a_3} - \frac{1}{a_1}\right) F(t)\ , \nonumber \\
  p_2(t) &=& \frac 14  \left( \frac{1}{a_3} + \frac{1}{a_1} - \frac{1}{a_2}\right) F(t) \ ,\\
  p_3(t) &=& \frac 14  \left( \frac{1}{a_1} + \frac{1}{a_2} - \frac{1}{a_3}\right) F(t) \ , \nonumber
\end{eqnarray}
and $p_0(t) = 1 - p_1(t) - p_2(t)-p_3(t)$. In particular, taking $a_1=a_2=a$ and $a_3=\infty$ one finds
\begin{equation}\label{}
  \tilde{\kappa}_1(s) =  \tilde{\kappa}_2(s) = \frac{-s \tilde{f}(s)}{a - \tilde{f}(s)} \ , \ \ \   \tilde{\kappa}_3(s) = 0 \ ,
\end{equation}
and hence
\begin{equation}\label{}
  \tilde{k}_1(s) =  \tilde{k}_2(s) = 0\ , \ \ \   \tilde{k}_3(s) = \frac 12  \frac{s \tilde{f}(s)}{a - \tilde{f}(s)}  \ ,
\end{equation}
gives rise to the legitimate memory kernel
\begin{equation}\label{k-3}
  K_t[\rho] = k_3(t) (\sigma_3 \rho \sigma_3 - \rho) ,
\end{equation}
with arbitrary $f(t)$ and $a>0$ satisfying additional condition
\begin{equation}\label{}
  0 \leq F(t) := \int_0^t f(\tau) d\tau \leq 2a ,
\end{equation}
for all $t \geq 0$. The corresponding solution reads
\begin{eqnarray}
  p_0(t) &=& 1 - \frac{1}{2a} F(t) \ , \nonumber  \\
  p_1(t) &=& p_2(t) = 0 \ , \\
  p_3(t) &=&  \frac{1}{2a} F(t)\ . \nonumber
\end{eqnarray}
This approach resembles very much the semi-Markov construction \cite{B-V,EPL}: for any $f(t) \geq 0$ satisfying $\int_0^\infty f(t)dt \leq 1$ the memory kernel (\ref{k-3}) with
\begin{equation}\label{}
  \tilde{k}_3(s) = \frac{s \tilde{f}(s) }{1-\tilde{f}(s) } ,
\end{equation}
gives rise to CPTP evolution. In this case one finds
\begin{eqnarray}
  p_0(t) &=& \frac 12 [ 1 + \lambda_1(t) ] \ , \nonumber  \\
  p_1(t) &=& p_2(t) = 0 \ , \\
  p_3(t) &=&  \frac 12 [ 1 - \lambda_1(t) ] \ , \nonumber
\end{eqnarray}
where
\begin{equation}\label{}
  \tilde{\lambda}_1(s) =  \tilde{\lambda}_2(s) = \frac{ \tilde{f}(s)+1}{ \tilde{f}(s)-1} \ .
\end{equation}
It is therefore clear that our approach goes beyond the semi-Markov construction.

Let us recall that Markovian semigroup generated by
\begin{equation}\label{}
  L[\rho] = \frac 12 \sum_{k=1}^3 \gamma_k [\sigma_k \rho \sigma_k - \rho] ,
\end{equation}
the corresponding Bloch equation reads
\begin{equation}\label{}
  \dot{x}_k(t) = - \frac{2}{T_k} x_k(t) \ ,
\end{equation}
where $x_k := {\rm tr}[\rho \sigma_k]$ and the relaxation times are defined via
\begin{equation}\label{}
  T_1 = \frac{1}{\gamma_2 + \gamma_3} \ , \ \   T_2 = \frac{1}{\gamma_3 + \gamma_1} \ , \ \   T_3 = \frac{1}{\gamma_1 + \gamma_2} \ .
\end{equation}
It is well known \cite{Alicki} that complete positivity is equivalent to the following set of conditions upon $T_k$:
\begin{eqnarray} \label{TTT}
  \frac{1}{T_1} + \frac{1}{T_2}  &\geq& \frac{1}{T_3} \ , \nonumber \\
  \frac{1}{T_2} + \frac{1}{T_3}  &\geq& \frac{1}{T_1} \ , \\
  \frac{1}{T_3} + \frac{1}{T_1}  &\geq& \frac{1}{T_2} \ , \nonumber
\end{eqnarray}
It is therefore clear that condition (\ref{aaa}) is an analogue of (\ref{TTT}). Note that condition (\ref{aaa}) means that there exist $b_1,b_2,b_3 >0 $ such that
\begin{eqnarray} \label{aaa-bbb}
  \frac 12 \frac{1}{a_1} &=& \frac{1}{b_2} + \frac{1}{b_3} \ , \nonumber \\
  \frac 12 \frac{1}{a_2} &=& \frac{1}{b_3} + \frac{1}{b_1} \ , \\
  \frac 12 \frac{1}{a_3} &=& \frac{1}{b_1} + \frac{1}{b_2} \ . \nonumber
\end{eqnarray}
Now, it terms of $b_1,b_2,b_3$ our result may be reformulated as follows

\begin{Corollary} For any $b_1,b_2,b_3 > 0$ and the function $f(t)$ satisfying
\begin{equation}\label{}
  0 \leq F(t) := \int_0^t f(\tau) d\tau \leq \left(\frac{1}{b_1} +  \frac{1}{b_2} + \frac{1}{b_3} \right)^{-1},
\end{equation}
and
\begin{equation}\label{}
  \lim_{s\rightarrow \infty} \tilde{f}(s) = 0 ,
\end{equation}
the memory kernel defined by
\begin{equation}\label{}
  \tilde{\kappa}_k(s) = -\frac{s \tilde{f}(s)}{a_k - \tilde{f}(s)}\ ,
\end{equation}
defines legitimate quantum evolution. Moreover one has
\begin{equation}\label{}
  p_k(t) = \frac{1}{b_k} F(t) \ ,
\end{equation}
and $p_0(1) = 1- p_1(t) - p_2(t) - p_3(t)$.
\end{Corollary}

Let us observe that it is very hard, in general, to invert formula (\ref{tilde-kappa}) to the time domain. Now, we provide a family of $W(s)$ which enables one to easily compute $\kappa_i(t)$ and have the memory kernel in  time domain.

\begin{Theorem} Let $W(s)$ be a polynomial
\begin{equation}\label{W}
   W(s) = (s+z_1)\ldots (s+z_n) ,
\end{equation}
with $z_i > 0$. If $a_1,a_2,a_2$ satisfy (\ref{aaa}) and
\begin{equation}\label{z-aaa}
  \prod_{i=1}^n z_i \geq \frac 14  \left( \frac{1}{a_1} + \frac{1}{a_2} + \frac{1}{a_3}  \right) ,
\end{equation}
then $\kappa_i(t)$ defined via (\ref{kapcm}) define a legitimate memory kernel.
\end{Theorem}
Proof: It is clear that it is enough to prove (\ref{s-0}).

\begin{Lemma}\label{lem} One has the following decomposition
\begin{gather}\label{wz1}
\frac{1}{s\prod_{i=1}^n(s+z_i)}=A\left(\frac{1}{s}-\sum_{i=1}^n\frac{\prod_{j=1}^{i-1}z_j}{\prod_{j=1}^i (s+z_j)}\right),
\end{gather}
where
\begin{equation}
A=\frac{1}{\prod_{i=1}^{n} z_i}.
\end{equation}
\end{Lemma}
For the proof see the Appendix. Now we show that condition (\ref{s-0}) holds. According to (\ref{wz1}) one has
\begin{eqnarray}\label{prop3}
&& \frac1s +  \tilde{\lambda}_1(s) + \tilde{\lambda}_2(s) + \tilde{\lambda}_3(s) \nonumber \\ &=& \frac{1}{s}\left( 4
- \left[  \frac{1}{a_1} + \frac{1}{a_2} + \frac{1}{a_3}\right] \frac{1}{W(s)}  \right) \nonumber \\ &=&
\frac 1s \left( 4 - \left[  \frac{1}{a_1} + \frac{1}{a_2} + \frac{1}{a_3}\right] \frac{1}{\prod_{i=1}^{n} z_i} \right)  \\
 & +& \left[  \frac{1}{a_1} + \frac{1}{a_2} + \frac{1}{a_3}\right] \frac{1}{\prod_{i=1}^{n} z_i} \sum_{j=1}^n\frac{\prod_{i=1}^{j-1}z_i}{\prod_{i=1}^j (s+z_i)} .\nonumber
\end{eqnarray}
A second term in (\ref{prop3}) is CM due to the fact that it is a sum of CM functions. Hence, if condition (\ref{z-aaa}) is satisfied then (\ref{s-0}) holds. \hfill $\Box$

Note, that
\begin{equation}\label{sss}
  \tilde{\kappa}_i(s) = - \frac{s}{a_kW(s)-1} = - \frac{1}{a_k}\frac{s}{(s-s_1) \ldots (s-s_m)} ,
\end{equation}
where $\{s_1,\ldots,s_m\}$ are the roots of the polynomial $(a_kW(s)-1)$. It is therefore clear that formula (\ref{sss}) may be easily inverted to the time domain.

\begin{Remark} Note, that $W(s)$ defined in (\ref{W}) implies that $\frac{1}{W(s)}$ is CM and hence $\frac 1s \frac{1}{W(s)}$ is CM as well.
\end{Remark}

\section{Checking for non-Markovianity}   \label{MAR}

Let us recall that according to \cite{BLP} the evolution represented by $\Lambda_t$ is non-Markovian
if the condition
\begin{equation}\label{blp}
  \frac{d}{dt} || \Lambda_t[\rho_1 - \rho_2]||_{\rm tr} \leq 0 \ ,
\end{equation}
is violated for some initial states $\rho_1$ and $\rho_2$.  One defines \cite{BLP} a well-known non-Markovianity measure
\begin{equation}\label{}
  \mathcal{N}_{\rm BLP}[\Lambda_t] = \sup_{\rho_1,\rho_2} \int  \frac{d}{dt} || \Lambda_t[\rho_1 - \rho_2]||_{\rm tr} \, dt\ ,
\end{equation}
where the integral is evaluated over the region where $\frac{d}{dt} || \Lambda_t[\rho_1 - \rho_2]||_{\rm tr} > 0$. Now, it has been proved \cite{PLA} that for random unitary qubit evolution if all eigenvalues $\lambda_k(t) \geq 0$, then (\ref{blp}) is equivalent to
\begin{equation}\label{}
  \frac{d}{dt} \lambda_k(t) \leq 0 \ ; \ \ \ k=1,2,3.
\end{equation}

\begin{Proposition} For $a_1,a_2,a_3$ satisfying (\ref{aaa}) and  $W(s)=\frac{1}{\tilde{f}(s)}$, where $\tilde{f}(s)$ is CM and
 \begin{equation}\label{int}
 \int_0^t f(\tau)\ud\tau\le a_{\mathrm{min}},\quad a_{\mathrm{min}}=\min\{a_1,a_2,a_3\},
 \end{equation}
the corresponding memory kernel gives rise to the dynamical map $\Lambda_t$ such that $ \mathcal{N}_{\rm BLP}[\Lambda_t]=0$.
\end{Proposition}
Proof: Let us observe that condition (\ref{int}) implies (\ref{aaa-0}). Indeed,  from (\ref{int}) one has
\begin{equation}
\left(\frac{1}{a_1}+\frac{1}{a_2}+\frac{1}{a_3}\right)\int_0^tf(\tau)\ud\tau\le3 ,
\end{equation}
and hence the condition (\ref{aaa-0}) follows. Now, observe that
$$\lambda_k(t)=1-\frac{1}{a_k}\int_0^t f(\tau)\ud\tau\ge0, $$
due to  (\ref{int}).  Hence it is sufficient to show that $\frac{d}{dt} \lambda_k(t) \leq 0$. It is clear $\frac{d}{dt} \lambda_k(t) \leq 0$ if and only if $1-s\tilde{\lambda}_k(s)$ is CM and hence taking into account (\ref{l-k}) it is equivalent to the requirement that $- \tilde{\kappa}_k(s) \tilde{\lambda}_k(s)$ is CM. One has therefore
\begin{equation}\label{}
  - \tilde{\kappa}_k(s) \tilde{\lambda}_k(s) = \frac{\tilde{f}(s)}{a_k },
\end{equation}
which ends the proof since $\tilde{f}(s)$ is CM and $a_k >0$  \hfill $\Box$

\begin{Remark} It $W(s) = (s+z_1)\ldots (s+z_n)$ with $z_k > 0$ and $a_1,a_2,a_3$ satisfying (\ref{aaa}) together with
\begin{equation}\label{z-123}
   \prod_{i=1}^n z_i \geq  \frac{1}{a_k}\ , \ \ \ k=1,2,3,
\end{equation}
then the corresponding dynamical map $\Lambda_t$ satisfies $ \mathcal{N}_{\rm BLP}[\Lambda_t]=0$.
\end{Remark}

\begin{Remark} It was shown \cite{Sab,Filip-PRA} that BLP condition (\ref{blp}) is equivalent to so-called P-divisibility. This means that
\begin{equation}\label{}
  \Lambda_t = V_{t,s} \Lambda_s,
\end{equation}
and for any $t > s$ the propagator $V_{t,s}$ is positive (but not necessarily completely positive).
\end{Remark}
Interestingly, our construction provides a class of legitimate random unitary qubit evolutions generated by the nontrivial memory kernel but still satisfying BLP condition (\ref{blp}), (cf. also \cite{Laura}). It is clear that to violate (\ref{blp}) one needs a more refined construction such that $\frac{1}{W(s)}$ is not CM but $\frac 1s \frac{1}{W(s)}$ is already CM. It deserves further analysis.

Consider now the question of CP-divisibility which is fully controlled by the local decoherence rates in (\ref{L}). 	
One may easily compute them in terms of $f(t)$:

\begin{eqnarray*}\label{gammas}
\gamma_1(t)&=&\frac{f(t)}{4}\left(\frac{-1}{a_1-F(t)}+\frac{1}{a_2-F(t)}+\frac{1}{a_3-F(t)}\right),\nonumber\\
\gamma_2(t)&=&\frac{f(t)}{4}\left(\frac{1}{a_1-F(t)}-\frac{1}{a_2-F(t)}+\frac{1}{a_3-F(t)}\right),\\
\gamma_3(t)&=&\frac{f(t)}{4}\left(\frac{1}{a_1-F(t)}+\frac{1}{a_2-F(t)}-\frac{1}{a_3-F(t)}\right).\nonumber
\end{eqnarray*} 
The dynamical map $\Lambda_t$ is CP-divisible iff $\gamma_k(t)\ge0$ for $k=1,2,3$. Let us assume that
\begin{equation}\label{<<}
 a_1 \leq a_2 \leq a_3 .   
\end{equation}

\begin{Proposition}
If $a_1,a_2,a_3$ and $f(t) \geq 0$ satisfy conditions (\ref{aaa-0}) and (\ref{aaa}) the corresponding memory kernel 
\begin{equation}
\tilde{\kappa}_k(s)=-\frac{s\tilde{f}(s)}{a_k-\tilde{f}(s)},
\end{equation}
leads to a CP-divisible dynamical map  iff 
 \begin{equation}  \label{a(aa)}
 F(t)\le a_1-\sqrt{(a_2-a_1)(a_3-a_1)} .
 \end{equation}
\end{Proposition}
Proof: Due to (\ref{<<}) it is sufficient to show that $\gamma_1(t)\geq 0$ which, for $f(t) \geq 0$,  is equivalent to 
\begin{equation}\label{ine}
   \frac{-1}{a_1-F(t)}+\frac{1}{a_2-F(t)}+\frac{1}{a_3-F(t)}\ge0. 
\end{equation}
Let us assume that $F(t)<a_1$, which means, that $\gamma_1(t)$ is not singular. 
Inequality (\ref{ine}) is satisfied iff 
$$   F(t) \in (-\infty,F_-] \cup [F_+,+\infty) $$
with
$$   F_\pm = a_1 \pm \sqrt{(a_2-a_1)(a_3-a_1)} . $$
Now, taking into account that $F(t) < a_1$ one finally proves (\ref{a(aa)}).   \hfill $\Box$

This Proposition shows that positivity of the function $f(t)$ is not sufficient for CP-divisibility. One needs an extra condition (\ref{a(aa)}) which involves not only $f(t)$ but $\{a_1,a_2,a_3\}$ as well.

\section{Examples}

\begin{Example}\label{ex1}
Consider the simplest case with a polynomial of degree one
\beq
 W(s)=  s + z ,
\eeq
with $z > 0$. One finds
\beq
\tilde{\kappa}_k(s) = -\frac{s}{a_k(s+z)-1} ,
\eeq
and the inverse Laplace transform gives
\beq
\kappa_k(t)= - \frac 1z \left( \delta(t) - \left[z - \frac{1}{a_k} \right]  e^{-[z - \frac{1}{a_k}] t} \right) .
\eeq
Note, that if $a_k = 1/z$, then the dynamics is purely local. One easily finds
\begin{equation}\label{}
  \lambda_k(t) = 1 - \frac{1}{za_k} (1-e^{-zt}) ,
\end{equation}
and finally the solution for $p_k(t)$ is defined by (\ref{pF}) with
\begin{equation}\label{}
  F(t) = \frac 1z (1 - e^{-zt}) \ .
\end{equation}

Note, that condition (\ref{z-aaa}) implies the following relation between $z$ and $a_1,a_2,a_3$:
\begin{equation}\label{}
  4z \geq \frac{1}{a_1} + \frac{1}{a_2} + \frac{1}{a_3}  ,
\end{equation}
which guarantees that $p_0(t) \geq 0$. In the symmetric case $a_1=a_2=a_3=a$ one finds $p_1(t) = p_2(t) = p_3(t) =: p(t)$ with
\begin{equation}\label{}
  p(t) =  \frac{1}{4za}  [ 1 - e^{-zt} ] ,
\end{equation}
and $p_0(t) = 1-3p(t)$ with $4za \geq 3$. One finds that asymptotically
\begin{equation}\label{}
  p_0(t) \rightarrow 1 - \frac{3}{4za}.
\end{equation}
Note that for $za > 1$ one has asymptotically $p_0(\infty) < 1/4$. This property cannot be reproduced within the local approach with regular generators $L_t$. Indeed, it follows from (\ref{p-H}) (see also \cite{PLA} for more details) that
\begin{equation}\label{}
  p_0(t) = \frac 14 [ 1 + \lambda_1(t) +\lambda_2(t) + \lambda_3(t)],
\end{equation}
and hence, using (\ref{lambda-123}), one finds
\begin{equation}\label{}
  p_0(t) \geq \frac 14.
\end{equation}
This example shows that local and memory kernel approaches may lead to essentially different evolutions.
\end{Example}


\begin{Example} Consider now the same polynomial $W(s) = s +z$ but let $z=2c > 0$. Moreover
\begin{equation}\label{}
  a_1=a_2 = \frac 1c \ , \ \ \ a_3 = \frac{1}{2c} .
\end{equation}
One finds
\begin{equation*}
\tilde{\kappa}_1(s)=\tilde{\kappa}_2(s)=-\frac{sc}{s+c},\quad \tilde{\kappa}_3(s)=-2c ,
\end{equation*}
and hence
\begin{equation*}\label{}
  \kappa_1(t) = \kappa_2(t) = - c \delta(t) + c^2  e^{-ct} \ , \ \kappa_3(t) = -2c \delta(t) .
\end{equation*}
Finally, one finds the following formula for the memory kernel
\begin{eqnarray}\label{}
  K_t[\rho] &=& \frac c2 \delta(t)[ \sigma_1 \rho \sigma_1 +  \sigma_2 \rho \sigma_2 - 2 \rho] \nonumber \\
  &-& \frac{c^2}{2} e^{-ct} [\sigma_3 \rho \sigma_3 - \rho] .
\end{eqnarray}
One has
\begin{equation*}
\lambda_1(t)=\lambda_2(t)=\frac{1}{2}\left(1+e^{-2ct}  \right),\quad\lambda_3(t)=e^{-2ct}.
\end{equation*}
Interestingly, this evolution reproduces time-local description with
\begin{equation}
\gamma_1(t)=\gamma_2(t)= \frac c2,\quad\gamma_3(t)=- \frac{c}{2} \tanh(ct).
\end{equation}
as discussed in \cite{Erika}. It was shown \cite{Filip-PRA} that $\Lambda_t$ is a convex combination of two Markovian semigroups $\Lambda^{(1)}_t$ and $\Lambda^{(2)}_t$ generated by
\begin{equation}\label{}
  L_k[\rho] = \frac c2 [ \sigma_k \rho \sigma_k - \rho] \ ; \ \ k=1,2,
\end{equation}
that is,
\begin{equation}\label{}
  \Lambda_t = \frac 12 \left( e^{tL_1} + e^{tL_2} \right) .
\end{equation}

This simple example shows that a convex combination of Markovian semigroups leads to a quantum evolution displaying essential memory effects.
\end{Example}

\begin{Example} Consider now a polynomial of degree two
\begin{equation}\label{}
  W(s) = (s+c_1)(s+c_2) ,
\end{equation}
with $c_2 > c_1 > 0$. Our construction gives rise to a legitimate memory kernel if condition (\ref{aaa}) holds and
\begin{equation}\label{}
  4 c_1 c_2 \geq \frac{1}{a_1} + \frac{1}{a_2} + \frac{1}{a_3} .
\end{equation}
One finds
\begin{eqnarray}\label{}
  \tilde{\kappa}_k(s) &=& -\frac{1}{a_k} \frac{s}{(s+c_1)(s+c_2) - \frac{1}{a_k} } \nonumber \\ &=& -\frac{1}{a_k} \frac{s}{(s+s_1)(s+s_2)} ,
\end{eqnarray}
with
$$ s_1 + s_2 = c_1+c_2\ , \ \ \ s_1 s_2 = c_1 c_2 - \frac{1}{a_k} \ . $$
Hence the solution has the form (\ref{pF}) with the function $F(t)$ given by
\begin{equation}\label{}
  F(t) = \frac{1}{c_2-c_1} \left( \frac{1}{c_1} [1- e^{-c_1 t}] -   \frac{1}{c_2} [1- e^{-c_2 t}] \right) .
\end{equation}

\end{Example}

\begin{Example}\label{ex3}
Let
\beq
 W(s)=s^2+\omega^2 .
\eeq
Note that $ \frac 1s \frac{1}{W(s)} $ is CM since
\begin{equation*}\label{}
  \frac 1s \frac{1}{W(s)} = \frac{1}{\omega} \frac 1s \left( \frac{\omega}{s^2+\omega^2}\right) ,
\end{equation*}
is the Laplace transform of $\int_0^t \sin(\omega \tau)d\tau$ which is positive for all $t \geq 0$. Condition (\ref{aaa-0}) implies
\begin{equation}\label{omi}
  \frac{1}{a_1} + \frac{1}{a_2} + \frac{1}{a_3} \leq 2 \omega^2 \ .
\end{equation}
The corresponding eigenvalues of the memory kernel read
\begin{equation}\label{}
  \kappa_i(t) = - \frac{1}{a_i} \cos\left( \sqrt{\omega^2 - \frac{1}{a_i}} \ t \right) ,
\end{equation}
for $\omega^2 \geq 1/a_i$,  and  
\begin{equation}\label{}
  \kappa_i(t) = - \frac{1}{a_i} \cosh\left( \sqrt{\frac{1}{a_i}-\omega^2} \ t \right) ,
\end{equation}
for $\omega^2 < 1/a_i$. Moreover one finds
\begin{equation}\label{}
  \lambda_k(t) = 1 + \frac{1}{a_k \omega^2} [ \cos(\omega t)  - 1 ] \ ,
\end{equation}
and hence
\begin{eqnarray} \label{}
  p_1(t) &=& \frac{1}{4\omega^2}  \left( \frac{1}{a_2} + \frac{1}{a_3} - \frac{1}{a_1}\right) [ 1-\cos(\omega t)]\ , \nonumber \\
  p_2(t) &=& \frac{1}{4\omega^2}  \left( \frac{1}{a_3} + \frac{1}{a_1} - \frac{1}{a_2}\right) [ 1-\cos(\omega t)] \ ,\\
  p_3(t) &=& \frac{1}{4\omega^2}  \left( \frac{1}{a_1} + \frac{1}{a_2} - \frac{1}{a_3}\right) [ 1-\cos(\omega t)] \ , \nonumber
\end{eqnarray}
together with $p_0(t) = 1 - p_1(t) - p_2(t)-p_3(t)$. In particular taking
\begin{equation}\label{}
  a_1=a_2=\frac{1}{\omega^2} , \ \ \ a_3 = \infty,
\end{equation}
one finds
\begin{equation}\label{}
   \kappa_1(t) =  \kappa_2(t) = - \omega^2\ , \ \ \  \kappa_3(t) = 0 \ ,
\end{equation}
and hence
\begin{equation}\label{}
   k_1(t) =  k_2(t) = 0\ , \ \ \  \kappa_3(t) =  \frac{\omega^2}{2} \ ,
\end{equation}
which proves that the constant (time independent)
\begin{equation}\label{}
  K_t[\rho] = \frac{k}{2} (\sigma_3 \rho \sigma_3 - \rho) ,
\end{equation}
provides a legitimate memory kernel for arbitrary $k = \omega^2 > 0$. Moreover one finds  for the local decoherence rates

\begin{eqnarray*}
&& \gamma_1(t) = \frac{\omega\sin(\omega t)}{4}\Big(\frac{-1}{a_1\omega^2-1+\cos(\omega t)} \nonumber \\ && + \frac{1}{a_2\omega^2-1+\cos(\omega t)} + \frac{1}{a_3\omega^2-1+\cos(\omega t)}\Big),
\end{eqnarray*}
and similarly for $\gamma_2(t)$ and $\gamma_3(t)$. Note, that if for some $k$ one has $a_k \omega^2 < 1$  then local decoherence rates are singular and hence in this case the non-local approach is more suitable.
\end{Example}


\section{Conclusions}

We analyzed random unitary evolution of a qubit within memory kernel approach. Our main result formulated in Theorem \ref{TH} allows to construct legitimate memory kernels leading to CPTP dynamical maps. The power of this method is based on the fact that 1) it allows to reconstruct well known examples of legitimate qubit evolution, 2) the structure of polynomials $W_k(s)$ enables one to perform the inverse Laplace transform and to find a formula for the kernel in the time domain. The mathematical analysis heavily uses the notion of completely monotone functions. These functions are not commonly used in theoretical physics and  knowledge of their properties is rather limited. There are no known effective methods allowing to check whether a given function is CM. We stress that Theorem \ref{TH} provides only a sufficient condition and further analysis is needed to cover physically interesting cases which do not fit the assumptions of the Theorem.  Interestingly, it turns out that the quantum evolution with a memory kernel generated by our approach gives rise to vanishing non-Markovianity measure based on the distinguishability  of quantum states \cite{BLP}. We also have shown when the corresponding dynamical map is CP-divisible. It shows that the evolution satisfying nonlocal master equation does not necessarily lead to a non-Markovian evolution.
It would be also interesting to analyze the relation between semi-Markov evolution and the one governed by our approach in more detail.

\section*{Acknowledgements}

We thank anonymous referee for valuable comments and Karol \.Zyczkowski for pointing out \cite{Karol}. This article was partially supported by the National Science Center project DEC-
2011/03/B/ST2/00136.

\section*{Appendix: proof of Lemma \ref{lem}}

Let us observe  that  (\ref{wz1}) may be represented in the following form
\begin{widetext}
\begin{equation}
\frac{1}{s\prod_{i=1}^n(s+z_i)}=A\frac{\prod_{i=1}^n(s+z_i)-s\left(\prod_{i=2}^n(s+z_i)+z_1\prod_{i=3}^n(s+z_i)+\ldots\prod_{j=1}^{n-2}z_j(s+z_n)+\prod_{j=1}^{n-1}z_j   \right)   }{s\prod_{i=1}^n(s+z_i)},
\end{equation}
\end{widetext}
therefore, to prove the Lemma it suffices to show that
\begin{widetext}
\begin{equation}\label{induction}
\prod_{i=1}^nz_i=\prod_{i=1}^n(s+z_i)-s\left(\prod_{i=2}^n(s+z_i)+z_1\prod_{i=3}^n(s+z_i)+\ldots\prod_{j=1}^{n-2}z_j(s+z_n)+\prod_{j=1}^{n-1}z_j   \right).
\end{equation}
\end{widetext}
We will prove this by induction. For $n=1$ it is clear that LHS=RHS=$z_1$. We assume that (\ref{induction}) is true for $n$ and prove  it is also true for $(n+1)$. LHS may be written as
\begin{equation}
\lhs=\prod_{i=1}^n z_i\cdot z_{n+1},
\end{equation}
while RHS reads
\begin{widetext}
\begin{eqnarray}
\rhs&=&\prod_{i=1}^n(s+z_i)(s+z_{n+1})-s\Big(\prod_{i=2}^n(s+z_i)(s+z_{n+1})+z_1\prod_{i=3}^n(s+z_i)(s+z_{n+1})+\ldots+\nonumber\\
&+&\prod_{j=1}^{n-2}z_j(s+z_n)(s+z_{n+1})+\prod_{j=1}^{n-1}z_j(s+z_{n+1})+\prod_{j=1}^{n-1}z_j\cdot z_n   \Big)=\nonumber\\
&=&(s+z_{n+1})\left(\prod_{i=1}^n(s+z_i)-s\left(\prod_{i=2}^n(s+z_i)+z_1\prod_{i=3}^n(s+z_i)+\ldots\prod_{j=1}^{n-2}z_j(s+z_n)+\prod_{j=1}^{n-1}z_j   \right)\right)-\nonumber\\
&-&s\prod_{j=1}^{n-1}z_j z_n=s\prod_{i=1}^nz_i+z_{n+1}\prod_{i=1}^n z_i-s\prod_{i=1}^n n z_i ,
\end{eqnarray}
\end{widetext}
which proves that RHS=LHS.  \hfill $\square$



\begin{thebibliography}{1} \bibliographystyle{plain}

\bibitem{Breuer} H.-P. Breuer and F. Petruccione,
{\em The Theory of Open Quantum Systems} (Oxford Univ. Press,
Oxford, 2007).

\bibitem{Weiss} U. Weiss, {\it Quantum Dissipative Systems}, (World
Scientific, Singapore, 2000).




\bibitem{GKS} V. Gorini, A. Kossakowski, and E. C. G. Sudarshan, J. Math. Phys.
{\bf 17}, 821 (1976).

\bibitem{L}  G. Lindblad, Comm. Math. Phys. {\bf 48}, 119
(1976).

\bibitem{Alicki} R. Alicki and K. Lendi, {\it Quantum Dynamical
Semigroups and Applications} (Springer, Berlin, 1987).


\bibitem{Gardiner} C.W. Gardiner and P. Zoller, {\it Quantum Noice},
Springer-Verlag, Berlin, 1999.


\bibitem{Plenio} M. B. Plenio and P. L. Knight, Rev. Mod. Phys. {\bf 70}, 101
(1998).

\bibitem{Car} H. J. Carmichael, {\em Statistical Methods in Quantum Optics I: Master Equations and Fokker-Plack Equations}, (Berlin: Springer 1999).



\bibitem{RIVAS} \'A. Rivas, S. F. Huelga, and M. B. Plenio,  Rep. Prog. Phys. {\bf 77}, 094001 (2014).

\bibitem{JPB} Special issue on Loss of Coherence and Memory Effects in
Quantum Dynamics, edited by F. Benatti, R. Floreanini, and
G. Scholes [J. Phys. B {\bf 45} (2012)].


\bibitem{Addis} C. Addis, B. Bylicka, D. Chru\'sci\'nski, S. Maniscalco, Phys. Rev. A {\bf 90}, 052103 (2014).




\bibitem{Wolf-Isert} M. M. Wolf, J. Eisert, T. S. Cubitt and J. I. Cirac, Phys. Rev. Lett. \textbf{101}, 150402 (2008).

\bibitem{RHP} \'A. Rivas, S.F. Huelga, and M.B. Plenio, Phys. Rev. Lett. {\bf 105}, 050403
(2010).

\bibitem{Sab} D. Chru\'sci\'nski and S. Maniscalco,  Phys. Rev. Lett. {\bf 112}, 120404 (2014).



\bibitem{NZ} S. Nakajima, Prog. Theor. Phys. {\bf 20}, 948 (1958); R.
Zwanzig, J. Chem. Phys. {\bf 33}, 1338 (1960).

\bibitem{NZ-inni} S. Chaturvedi and J. Shibata, Z. Phys. B {\bf 35}, 297 (1979); N. H. F. Shibata and Y. Takahashi,
J. Stat. Phys. {\bf 17}, 171 (1977).


\bibitem{Fabio} F. Benatti and R. Floreanini, Mod. Phys. Lett. A {\bf 12}, 1465 (1997).

\bibitem{Hangii} P. H\"anggi and H. Thomas, Z. Phys B: Condens. Matter {\bf 26}, 85 (1977); H. Grabert, P. Talkner and P. H\"anggi,  Z. Phys B: Condens. Matter {\bf 26}, 389 (1977); A. Fuliński and W.J. Kramarczyk, Physica {\bf 39}, 575 (1968).

\bibitem{PRL-2010} D. Chru\'sci\'nski and A. Kossakowski, Phys. Rev. Lett. {\bf
104}, 070406 (2010).











\bibitem{bar} S. M. Barnett and S. Stenholm, Phys. Rev. A {\bf 64}, 033808 (2001).

\bibitem{Budini} A. A. Budini, Phys. Rev. A {\bf 69}, 042107 (2004).

\bibitem{Wilke}   J. Wilkie,  Phys. Rev. E {\bf 62},  8808 (2000); J. Wilkie and Yin Mei
Wong, J. Phys. A: Math. Theor. {\bf 42}, 015006 (2009);

\bibitem{B-V} H.-P. Breuer and B. Vacchini, Phys.
Rev. Lett. {\bf 101} (2008) 140402; Phys. Rev. E {\bf 79}, 041147
(2009).


\bibitem{Wod} S. Daffer, K. W\'odkiewicz, J.D. Cresser, and J.K. McIver,
Phys. Rev. A {\bf 70}, 010304 (2004).

\bibitem{Lidar}  A. Shabani and D.A. Lidar, Phys. Rev. A {\bf 71}, 020101(R) (2005).


\bibitem{Maniscalco} S. Maniscalco, Phys. Rev. A {\bf 72}, 024103 (2005); S. Maniscalco and F. Petruccione, Phys.
Rev. A {\bf 73}, 012111 (2006).


\bibitem{AK} A. Kossakowski and R. Rebolledo, Open Syst. Inf. Dyn. {\bf 14}, 265 (2007); \textit{ibid.}
{\bf 16}, 259 (2009).

\bibitem{EPL} D. Chru\'sci\'nski and A. Kossakowski, EPL {\bf 97}, 20005 (2012).

\bibitem{Bassano} B. Vacchini,  Phys. Rev. A {\bf 87}, 030101(R) (2013).

\bibitem{Laura}L. Mazzola, E.-M. Laine, H.-P. Breuer, S. Maniscalco, and J. Piilo, Phys. Rev. A {\bf81}, 062120 (2010).












\bibitem{BLP} H.-P. Breuer, E.-M. Laine, J. Piilo, Phys. Rev. Lett. {\bf 103},
210401 (2009).













\bibitem{Ad} K. M. R. Audenaert and S. Scheel,  New J. Phys. {\bf 10},  023011 (2008).

\bibitem{Karol} In \cite{Alicki} random unitary evolution is called {\em dynamics for a system in random external fields}. In this paper we use {\em random unitary channel/evolution} following e.g. \cite{Ad}.  

\bibitem{PLA} D. Chru\'sci\'nski and F. A. Wudarski, Phys. Lett. A {\bf377}, 1425 (2013).

\bibitem{Erika}M. J. W. Hall,  J. D. Cresser, L. Li and E. Andersson,  Phys. Rev. A {\bf 89}, 042120 (2014).

\bibitem{Filip-PRA} D. Chru\'sci\'nski, and F. A. Wudarski,  Phys. Rev. A {\bf91}, 012104 (2015).

\bibitem{CM} K. S. Miller, and S. G. Samko, Integr. Transf. and Spec. Funct. {\bf12}, No 4, 389-402, (2001)








\end{thebibliography}
\end{document}